\documentstyle[pre,aps]{revtex}

\begin{document}

\draft

\title{Critical Indices as Limits of Control Functions}

\author{V.I. Yukalov and S. Gluzman}

\address{International Centre of Condensed Matter Physics \\
University of Brasilia, CP 04513, Brasilia, DF 70919--970, Brazil}

\maketitle

\begin{abstract}

A variant of self--similar approximation theory is suggested, 
permitting an easy and accurate summation of divergent series consisting 
of only a few terms. The method is based on a power--law algebraic 
transformation, whose powers play the role of control functions governing 
the fastest convergence of the renormalized series. A striking relation 
between the theory of critical phenomena and optimal control theory is 
discovered: The critical indices are found to be directly related to 
limits of control functions at critical points. The method is applied to 
calculating the critical indices for several difficult problems.
The results are in very good agreement with accurate numerical data.

\end{abstract}

\vspace{1cm}


The standard, and in many cases the sole, tool of theoretical physics is 
perturbation theory giving physical quantities in the form of power 
series. However, such perturbative series are usually divergent. If a 
sufficiently large number of terms of a divergent series are known, one 
can invoke resummation techniques. In theoretical physics, the most 
common resummation technique is Pad\'e summation [1]. Unfortunately, 
Pad\'e approximants and related techniques are often not able to produce 
sufficiently accurate results if only very few terms of a divergent 
series are available. Thus, one is frequently confronted with the 
annoying problem of extracting a sufficient amount of information from a 
divergent series even if only a few terms are known. An approach [2-6], 
called the self--similar approximation theory, for solving this problem was 
developed, combining the renormalization--group techniques [2-4] with the 
methods of dynamical theory and optimal control theory [5,6]. One of the main 
ideas of this approach [2-6] is to introduce control functions governing the 
fastest convergence of approximation sequences [7]. It was shown recently [8] 
that an effective way of introducing control functions is by means of a 
power--law algebraic transform with powers playing the role of these
control functions. Mathematical details of such an algebraic 
self--similar renormalization were expounded in Ref.[8]. 

In this paper we show that the approach can be extended to such a difficult
problem as the calculation of critical indices for complex physical 
models, for which only a couple of perturbative terms can be obtained. A 
very interesting relation between control functions and critical indices 
is discovered. We illustrate the method by calculating the critical 
indices for several quite complicated systems. The results are in very 
good agreement with numerical data available.

To make this paper self--consistent, we outline, first, the scheme of the 
method to be used. The complete mathematical foundation can be found in 
earlier publications [2-8]. Since the most elaborate presentation of the 
approach [5,6,8] relies on the language of dynamical theory and control 
theory, it would probably be helpful to give some additional references 
on dynamical theory [9] and optimal control theory [10] where a reader
could find more details on the mathematical language and techniques used 
in the approach.

Assume that we are interested in a function $\varphi(x)$ of a real variable 
$x$. Let perturbation theory give for this function perturbative 
approximations $p_k(x)$ with $k=0,1,2,\ldots$ enumerating the 
approximation order. Define the algebraic transform $F_k(x,c) = x^cp_k(x)$. 
This transform changes the powers of the series $p_k(x)$ changing by this the 
convergence properties of the latter. As a result, the approximation order 
effectively increases from $k$ to $k+c$. The inverse transform is 
$p_k(x) =x^{-c}F_k(x,c)$. Define the expansion function $x=x(f,c)$ by the 
equation $F_0(x,c) = f$, where $F_0$ is the first available approximation
and $f$ is a new variable. Substituting $x(f,c)$ back to $F_k$, we get 
$y_k(f,c) =F_k(x(f,c),c)$. The transformation inverse to the latter 
reads $F_k(x,c) = y_k(F_0(x,c),c)$.

Consider the family $\{ y_k\}$ as a dynamical system in discrete time. As far 
as the trajectory $\{ y_k(f,c)\}$ of this 
dynamical system is, by construction, bijective to the approximation sequence 
$\{ F_k(x,c)\}$, this system can be called the approximation cascade. 
Embed the discrete sequence $\{ y_k(f,c)\}$ into a continuous sequence 
$\{ y(\tau,f,c)\}$ with $\tau\in [0,\infty)$. 
Thence, the family $\{ y(\tau,\ldots):\;\tau\in[0,\infty)\}$ composes a 
dynamical system with continuous time, whose trajectory passes through all 
points of the approximation cascade trajectory. Because of this, such a system 
can be called the approximation flow. 

The evolution equation for a flow can be presented in the functional form 
$y(\tau+\tau',f,c)=y(\tau,y(\tau',f,c),c)$. We call this equation the 
{\it self--similarity} relation. Since the property of 
self--similarity is the central concept of our approach, let us remind 
where this notion is originated from. First of all, recall that this 
property is common for any autonomous dynamical system, reflecting the 
group property of motion [9]. Such relations appear in the 
renormalization--group approach of quantum field theory [11]. Recently, 
the concept of self--similarity was widely employed in the context of 
fractals. In all the cases, self--similarity is the group property of a 
function conserving its form under the change of its variable. In particular 
cases, this variable can be time, as for dynamical systems, momentum, as for 
field theory, or space scale, as for fractals. In our case, such a variable is
the approximation number, playing the role of time, and the motion occurs in 
the space of approximations, where self--similarity is a necessary condition 
for convergence [3,12]. The evolution equation for the approximation flow 
can be rewritten in the differential form and then integrated over time 
between $k$ and some $k^*$. 
The point $k^*$ is to be chosen so that to provide the best approximation
$F_{k+1}^*(x,c) = y(k^*,F_0(x,c),c)$ for the minimal time $k^*-k$. 
The cascade velocity $v_k(y,c)$ in the vicinity of the time moment $k$
may be presented by the Euler discretization of the flow velocity giving 
$v_k(f,c) = V_k(x(f,c),c)$, with $V_k(x,c) = F_{k+1}(x,c) - F_k(x,c)$.
The integral form of the evolution equation is
\begin{equation}
\int_{F_k}^{F_{k+1}^*}\frac{df}{v_k(f,c)} =k^*-k ,
\end{equation}
where $F_k=F_k(x,c)$ and $F_{k+1}^*=F_{k+1}^*(x,c)$. The 
approximation $F_{k+1}^*$ must be reached during the minimal time. 
When no additional constraints are imposed, then the minimal time 
corresponds, evidently, to one step: $k^*=k+1$. Finding 
$F_k^*(x,c)$ from (1), and using the inverse transform, we come to the 
self--similar approximation $p_k^*(x,c) = x^{-c}F_k^*(x,c)$.

Now, by means of the substitution $c\rightarrow c_k$, we have to 
introduce control functions $c_k$ which would govern the convergence 
of the sequence $\{ p_k^*(x,c_k)\}$. For the latter, following the 
standard procedure [2-6] and similar to the steps described above, we may 
construct an approximation cascade with the cascade velocity
\begin{equation}
V_k^*(x,c_k) =p_{k+1}^*(x,c_k) - p_k^*(x,c_k) +(c_{k+1}-c_k)
\frac{\partial}{\partial c_k} p_k^*(x,c_k) ,
\end{equation}
which is the Euler discretization taking into account the 
finite--difference variation due to the varying control function. 
Convergence of an approximation sequence is, in the language of dynamical 
theory, the same as the existence of an attracting fixed point for the 
corresponding approximation cascade. If the cascade trajectory tends to a 
fixed point, this means that $V_k(x,c_k)\rightarrow 0$ as 
$k\rightarrow\infty$. In practice, we cannot, of course, reach the limit 
$k\rightarrow\infty$, and have to stop at a finite $k$. Then, the 
condition to be as close to a fixed point as possible is the minimum of the 
velocity:
\begin{equation}
|V_k^*(x,c_k(x)|=\min_{c}|V_k^*(x,c)| .
\end{equation} 
To make Eq.(3) simpler, we notice [4-6] that in the vicinity of a stable 
fixed point we have $p_{k+1}^*\approx p_k^*, \; c_{k+1}\approx c_k$, and 
$\partial p_k^*/\partial c_k\approx 0$. Consequently, the last term in 
the velocity (2) is of the second order of smallness, as compared to the 
difference $p_{k+1}^*-p_k^*$. Therefore the practical way of using Eq.(3) 
is to start from the first--order term considering the minimal difference 
condition [7], that is, $\min_{c}|p_{k+1}^*(x,c)-p_k^*(x,c)|$, and, only 
if this has several solutions, then one has to chose that one which 
minimizes the 
second--order term. Note that the fixed point equations as in Eq.(3) have 
been used in Ref.[13].

After the control functions are found from the minimal--velocity 
condition (3), we substitute them into $p_k^*$ and obtain the final 
expression $f_k^*(x) = p_k^*(x,c_k(x))$
for the self--similar approximation of the sought function. To check 
whether the obtained sequence $\{ f_k^*(x)\}$ converges, we have to 
analyse whether the corresponding mapping is contracting. The mapping related 
to the sequence $\{ f_k^*\}$ is constructed in the standard way [2-8] and the 
contraction, or stability, is analysed by calculating the mapping 
multipliers.

Now we shall illustrate by several examples how our method of algebraic 
self--similar renormalization works. We show that we can describe the 
critical behaviour well, starting from virial--type expansions containing 
only second--order terms and derived for a region being far from the critical
point. We would like to stress that the general approach 
formulated above naturally suits the complicated and unsolved problem of 
deriving the critical indices from very short virial--type series and 
that the control functions introduced by means of the algebraic
transformation happen to be directly related to critical indices. Thus, 
it turns out surprisingly that {\it critical indices} can be considered 
as {\it physical analogs} of such mathematical objects as {\it control 
functions}.

Begin with polymer coils. For these one usually 
considers the expansion factor $\alpha(z)$ as a function of the 
excluded--volume variable $z\equiv B\sqrt{N}(3/2\pi l^2)^{3/2}$, in 
which $B$ is an effective binary cluster integral; $N$, the number 
of bonds of the length $l$ each [14-16]. For the region of 
$0\leq z\ll 1$, corresponding to the poor-solvent or weak--coupling case, 
perturbation theory in powers of $z$ gives [14-17] the expansion
$\alpha^2(z)\simeq 1 + a_1z + a_2z^2$. On the other hand, in the 
good--solvent or strong--coupling case, when $z\gg 1$, the polymer 
swelling is described by the law $\alpha^2(z)\simeq Az^{2(2\nu - 1)}$,
in which $\nu\geq 1/2$ is a critical index. 

\vspace{2mm}

(i) {\it 2D Polymer Coil}. In the two--dimensional case, the coefficients 
in the virial--type perturbative expansion [15] are: $a_1=0.5$ and 
$a_2=-0.12154525$. Starting from the first--order and second--order 
perturbative expressions $p_k(z)$ for $\alpha^2(z)$, following the 
prescribed way [8], as the self--similar approximations $p_k^*$, when 
$z\rightarrow\infty$, we obtain $p_k^*(z,c)\simeq A_kz^{-c}$,
where $A_1=\left (\frac{c}{-a_1}\right )^c$ and $A_2 = 
a_1^{2+c}\left (\frac{1+c}{-a_2}\right )^{1+c}$. Comparing this with 
$\alpha^2(z)$, as $z\rightarrow\infty$, we get $c=2(1-2\nu)$. 
Here $c<0$ is the limit, as $z\rightarrow\infty$, of the control 
function $c(z)$. Thus, we have the relation
\begin{equation}
\nu \equiv \frac{1}{2} 
+\frac{1}{2}\lim_{z\rightarrow\infty}\frac{\ln\alpha(z)}{\ln z} =
\frac{1}{2} -\frac{1}{4}\lim_{z\rightarrow\infty}c(z) 
\end{equation}
between the critical index $\nu$ and the limit of the control
function $c(z)$. From the fixed--point equation in the form of 
the minimal velocity condition (3), we find $c=-1$, which gives 
$\nu=0.75$. This is the {\it exact} value for the critical index [17]. 
Note that the Flory [18] mean--field formula $\nu=3/(2+D)$, valid for 
$D<4$, also reproduces this exact value. The reason why the mean--field 
approach can become exact can be understood if one notices [17] that the 
$2D$ polymer corresponding, within the framework of the $O(n)$--model, 
to $n\rightarrow 0$ belongs to the same line in the $n-D$ plane as 
the Gaussian polymer with $n=-2$. Along this line, the mean--field 
gives exact critical indices [19]. 

\vspace{2mm}

(ii) {\it 3D Polymer Coil}. In the three-dimensional case, the coefficients 
of the perturbative expansion [14-16] are: $a_1 =4/3$ and
$a_2=28\pi/27-16/3$. Now it is more convenient, for the stability of 
the procedure,  to invert $\alpha^2$ and to study 
$\alpha^{-2}(z)\simeq 1 + b_1z + b_2z^2$, where $b_1=-4/3$ and 
$b_2=3.853$. Repeating the same steps as for the 
$2$--dimensional case, only with $b_k$ instead of $a_k$, we 
get $c=2(2\nu-1)$. From the minimal velocity condition we find 
$\;c=0.396\;$, which yields $\nu=0.599$. This is close to the Flory [18] 
and Edwards [20] mean--field result $\nu=0.6$, as well as to the 
renormalization group calculations [21-24] giving $\nu\cong 0.589$. For 
the critical amplitude we find $A\cong 1.62$, which is also close to 
the renormalization group results [21-24] varying in the interval 
$1.53\leq A\leq 1.75$. 

The problem of the static conductivity of binary inhomogeneous materials 
is a well--known example of a rather complicated problem. Here one considers 
the  resistivity $\rho(x)$ and conductivity $\sigma(x)$ as functions of the 
concentration $x$ of the superconducting or conducting bonds or sites in a 
disordered system [25,26]. The critical behaviour is characterized 
by the so--called superconductivity, $s$, and conductivity, $t$, 
exponents describing the power laws of resistivity $\rho$ below the 
percolation threshold $x_c$ and of conductivity $\sigma$ above 
this threshold. The mean--field--type approximation for these disordered 
systems is the effective--medium approximation [25,27,28] which gives $s=t=1$ 
independently on space dimensionality. This is in sharp disagreement with 
the accurate numerical data [26] obtained by means of extensive 
transfer--matrix computer calculations. The position--space renormalization 
group [29-33] yields the results being also in a poor agreement with the 
numerical data [26]. 

\vspace{2mm}

(iii) {\it 3D Disordered Superconductor}. Below the percolation threshold, 
the effective--medium expression for the resistivity of bond--disordered 
systems [26-28] gives $\rho(x) = 1-3x$. From here we have for the 
self--similar renormalized resistivity $\rho_1^*(x,c) = c^c/(c+3x)^c$. 
This immediately gives us the relation 
\begin{equation}
s \equiv \lim_{x\rightarrow x_c}\frac{\ln\rho(x)}{\ln(x_c-x)} =
-\lim_{x\rightarrow x_c}c(x), \qquad x\leq x_c ,
\end{equation}
between the superconductivity exponent and the control function. From the 
equation $\rho(x_c)=0$ we find another relation, $c=-3x_c$. With the 
threshold [26] $x_c=0.2488$, we obtain $c=-0.746$. Then from Eq.(5) we get 
$s=0.746$, which is close to the numerical value $s=0.73\pm 0.01$ from 
Ref.[26]. 

\vspace{2mm}

(iv) {\it 2D Disordered Conductor}. Dealing with disordered conductors, 
it is convenient to exploit the fact that for them the effective--medium 
approximation becomes asymptotically exact in the low--defect--concentration 
limit, that is, in the pure--conductor limit $x\rightarrow 1$ 
[25-28]. Then, perturbation theory in powers of $x-1$ gives [34] for 
the two--dimensional case $\sigma(x) \simeq 1 -\pi(1-x) + 1.28588(1-x)^2$. 
Following the same procedure of the algebraic self--similar 
renormalization, as is described above, we find the relation
\begin{equation}
t \equiv \lim_{x\rightarrow x_c}\frac{\ln\sigma(x)}{\ln(x-x_c)} 
= -\lim_{x\rightarrow x_c} c(x) , \qquad x\geq x_c , 
\end{equation}
between the conductivity exponent and the control function. The latter is 
given by the minimal velocity criterion (3). The condition 
$\sigma_k^*(x_c)=0$ defines the threshold $x_c=0.5905$ which is close to 
the numerical value [26] being $x_c=0.5927$. From Eq.(6) we get $t=1.287$, 
which is again in a good agreement with the numerical value [26] being 
$t=1.299\pm 0.002$.

\vspace{2mm}

(v) {\it 3D Disordered Conductor}. For  the three--dimensional cubic 
lattice, perturbation theory gives [25] the expansion
$\sigma(x)\simeq 1 -2.52(1-x)+1.52(1-x)^2$. From the threshold 
equation $\sigma_k^*(x_c)=0$ we have $x_c=0.255$, which is close to 
the numerical result [26] for $x_c=0.3116$. The procedure of 
self--similar renormalization is the same as for the two--dimensional case,
yielding the same relation (6). For the conductivity index we find 
$t=1.88$ which practically coincides with the numerical result [26] 
giving $t=1.9\pm 0.1$.

To emphasize the generality of our method which works for perturbative 
series in powers of arbitrary expansion parameters, let us analyse one 
more nontrivial model.

(vi) {\it (2+1)D Ising Model}. Consider the so--called 
$(2+1)$--dimensional Ising model given by the hamiltonian 
$H=\sum_i(1-\sigma_i^z)-x\sum_{<ij>}\sigma_i^x\sigma_j^x 
-h\sum_i\sigma_i^x$, where $i$ and $j$ enumerate sites on the 
two--dimensional square lattice, $\langle ij\rangle$ denotes 
nearest--neighbour pairs, $\sigma_i^\alpha$ are the Pauli matrices, $x$ 
corresponds to the dimensionless inverse temperature in the Euclidian 
formulation, and $h$ is the magnetic field variable. The 
high--temperature expansion [35] for the susceptibility at zero field 
yields $\chi\simeq  1+4x+13.5x^2$, or for the inverse quantity $\chi^{-1} 
\simeq 1 -4x+2.5x^2$. Following again the same procedure as above, we 
find the relation
\begin{equation}
\gamma\equiv -\lim_{x\rightarrow x_c}\frac{\ln\chi}{\ln(x_c-x)} =
\lim_{x\rightarrow x_c}c(x), \qquad x\leq x_c ,
\end{equation} between the critical index $\gamma$ and the control 
function $c(x)$. For the critical point we get $x_c=0.308$ and for the 
index, $\gamma=1.232$. This is in good agreement with the Pad\'e 
summation of high--temperature series containing the terms up to 
sixteenth order [35], which gives $x_c=0.3285$ and $\gamma=1.245\pm 0.005$.

In conclusion, we have advanced here a new method allowing an easy and 
accurate summation of divergent series containing only a few terms. A very 
interesting fact has been discovered that the critical indices are directly 
related to control functions governing the convergence of renormalized series 
and providing the stability of calculational procedure. The calculated values 
for the critical indices are in very good agreement with numerical data 
and exact results when the latter are available. The intriguing relation
between the critical indices and control functions permits us to conjecture
that the former play the same role in nature as the latter in theory: 
{\it Nature choses such critical indices that provide the stability of 
physical phenomena in the vicinity of critical points}.


\begin{references}

\bibitem{1} G.A. Baker, Jr., and P. Graves--Morris, {\it Pad\'e Approximants}
(Cambridge Univ., Cambridge, 1996).

\bibitem{2} V.I. Yukalov, Phys. Rev. A {\bf 42}, 3324 (1990).

\bibitem{3} V.I. Yukalov, J. Math. Phys. {\bf 32}, 1235 (1991).

\bibitem{4} V.I. Yukalov, J. Math. Phys. {\bf 33}, 3994 (1992).

\bibitem{5} V.I. Yukalov and E.P. Yukalova, Physica A {\bf 206}, 553 (1994).

\bibitem{6} V.I. Yukalov and E.P. Yukalova, Physica A {\bf 225}, 336 (1996).

\bibitem{7} V.I. Yukalov, Mosc. Univ. Phys. Bull. {\bf 31}, 10 (1976).

\bibitem{8} S. Gluzman and V.I. Yukalov, Phys. Rev. E {\bf 55}, 3983 (1997).

\bibitem{9} P.A. Cook, {\it Nonlinear Dynamical Systems} (Prentice Hall, 
New York, 1994).

\bibitem{10} H. Tolle, {\it Optimization Methods} (Springer, New York, 1975).

\bibitem{11} N.N. Bogolubov and D.V. Shirkov, {\it Quantum Fields} (Benjamin,
London, 1983).

\bibitem{12} V.I. Yukalov, Physica A {\bf 167}, 833 (1990).

\bibitem{13} V.I. Yukalov and E.P. Yukalova, Can. J. Phys. {\bf 71}, 537 
(1993).

\bibitem{14} H. Yamakawa, {\it Modern Theory of Polymer Solutions} (Harper 
and Row, New York, 1971).

\bibitem{15} M. Muthukumar and B.G. Nickel, J. Chem. Phys. {\bf 80}, 5839 
(1984).

\bibitem{16} A.J. Barrett and C. Domb, J. Stat. Phys. {\bf 77}, 491 (1994).

\bibitem{17} B. Nienhis, Phys. Rev. Lett. {\bf 49}, 1062 (1982).

\bibitem{18} P.J. Flory, in {\it Principles of Polymer Chemistry} 
(Cornell Univ., Ithaca, 1971), Chap. 12.

\bibitem{19} V.I. Yukalov and A.S. Shumovsky, {\it Lectures on Phase 
Transitions} (World Scientific, Singapore, 1990).

\bibitem{20} S.P. Edwards, Proc. Phys. Soc. London {\bf 85}, 613 (1965).

\bibitem{21} J. Des Cloizeaux, R. Conte, and G. Jannink, J. Phys. Lett. 
{\bf 46}, L 595 (1985).

\bibitem{22} M. Muthukumar and B.G. Nickel, J. Chem. Phys. {\bf 86}, 460 
(1987).

\bibitem{23} K.F. Freed, {\it Renormalization Group Theory of 
Macromolecules} (Wiley, New York, 1987).

\bibitem{24} J.C. Le Guillou and J. Zinn--Justin, J. Phys. France {\bf 50}, 
1365 (1989).

\bibitem{25} S. Kirkpatrick, Rev. Mod. Phys. {\bf 45}, 574 (1973).

\bibitem{26} J.P. Clerc, G. Giraud, J.M. Laugier, and J.M. Luck, Adv. 
Phys. {\bf 39}, 191 (1990).

\bibitem{27} B.P. Watson and P.L. Leath, Phys. Rev. B {\bf 11}, 4893 (1974).

\bibitem{28} J. Bernasconi and H.J. Wiesmann, Phys. Rev. B {\bf 13}, 1131 
(1976).

\bibitem{29} R.B. Stinchcombe and B.P. Watson, J. Phys. C {\bf 9}, 3221 
(1976).

\bibitem{30} J.P. Straley, J. Phys. C {\bf 10}, 3009 (1997).

\bibitem{31} J. Bernasconi, Phys. Rev. B {\bf 18}, 2185 (1978).

\bibitem{32} J.M. Luck, J. Phys. A {\bf 18}, 2061 (1985).

\bibitem{33} M. Sahimi, B.D. Hudges, L.E. Scriven, and H.T. Davis, Phys. 
Rev. B {\bf 28}, 307 (1983).

\bibitem{34} T.M. Nieuwenhuizen, P.F. van Velthoven, and M.H. Ernst, 
Phys. Rev. Lett. {\bf 57}, 2477 (1986).

\bibitem{35} H.X. He, C.J. Hamer, and J. Oitmaa, J. Phys. A {\bf 23}, 
1775 (1990).

\end{references}
\end{document}